\documentclass[useAMS,usenatbib,usegraphicx]{mn2e}
\usepackage{times}
\usepackage{amsmath}
\usepackage{amssymb}

\input{psfig.sty}

\newcommand{\AIPS}{{$\cal AIPS\/$}}

\def\jonezero{$J$=1$\rightarrow$0}
\def\jthreetwo{$J$=3$\rightarrow$2}

\def\gs{\mathrel{\raise0.35ex\hbox{$\scriptstyle >$}\kern-0.6em
\lower0.40ex\hbox{{$\scriptstyle \sim$}}}}
\def\ls{\mathrel{\raise0.35ex\hbox{$\scriptstyle <$}\kern-0.6em
\lower0.40ex\hbox{{$\scriptstyle \sim$}}}}
\def\m@th{\mathsurround=0pt }
\def\eqalign#1{\null\,\vcenter{\openup1\jot \m@th
 \ialign{\strut\hfil$\displaystyle{##}$&$\displaystyle{{}##}$\hfil
 \crcr#1\crcr}}\,}

\title[Locating the starburst in the SCUBA galaxy, SMM\,J02399$-$0136]
      {Gas, dust and stars in the SCUBA galaxy, SMM\,J02399$-$0136:
      the EVLA reveals a colossal galactic nursery}

\author[Ivison et al.]
       {R.\,J.\ Ivison,$^{\! 1,2}$
Ian Smail,$^{\! 3}$
P.\,P.\ Papadopoulos,$^{\! 4}$
I.\ Wold,$^{\! 5}$
J.\ Richard,$^{\! 3}$ 
A.\,M.\ Swinbank,$^{\! 3}$ \and
J.-P.\ Kneib$^6$ and
F.\,N.\ Owen$^7$
 \vspace*{1mm}\\
$^1$ UK Astronomy Technology Centre, Science and Technology Facilities Council,
     Royal Observatory, Blackford Hill, Edinburgh EH9 3HJ\\
$^2$ Institute for Astronomy, University of Edinburgh, Blackford Hill,
     Edinburgh EH9 3HJ\\
$^3$ Institute for Computational Cosmology, Durham University,
     South Road, Durham DH1 3LE\\
$^4$ Argelander-Instit\"ut f\"ur Astronomie, Auf dem H\"ugel 71,
     D-53121, Germany\\
$^5$ Dept of Astronomy, University of Wisconsin-Madison, 475 N.\ Charter St.,
     Madison, WI\,53706, USA\\
$^6$ Laboratoire d'Astrophysique de Marseille, OAMP, CNRS-Universit\'e
     Aix-Marseille, 38 rue Fr\'ed\'eric Joliot-Curie, 13388 Marseille
     Cedex 13, France\\
$^7$ National Radio Astronomy Observatory, P.O.\ Box 0, Socorro, NM\,87801, USA
}

\date{Accepted ... ; Received ... ; in original form ...}

\pagerange{000--000}

\begin{document}

\maketitle

\begin{abstract}
We present new multi-wavelength observations of the first
submillimetre-selected galaxy (SMG), SMM\,J02399$-$0136 at
$z=2.8$. These observations include mapping of the CO \jonezero\
emission using elements of the Expanded Very Large Array, as well as
high-resolution 1.4-GHz imaging and optical/infrared (IR) data from
the Very Large Array, {\it Hubble Space Telescope}, {\it Spitzer} and
Keck-{\sc i}.  Together these new data provide fundamental insights
into the mass and distribution of stellar, gas and dust distribution
within this archetypal SMG. The CO \jonezero\ emission, with its
minimal excitation and density requirements, traces the bulk of the
metal-rich molecular gas, and reveals a molecular gas mass of
$\sim$10$^{11}$\,M$_{\odot}$, extending over approximately 5\,arcsec
($\sim$25\,kpc in the source plane), although there is tentative
evidence that it may be significantly larger. Our data suggest that
three or more distinct structures are encompassed by this molecular
gas reservoir, including the broad-absorption-line (BAL) quasar from
which the redshift of the SMG was initially determined.  In
particular, the new rest-frame near-IR observations identify a
massive, obscured, starburst which is coincident with a previously
known Ly$\alpha$ cloud.  This starburst dominates the far-IR emission
from the system and requires a re-assessment of previous claims that
the gas reservoir resides in a massive, extended disk around the BAL
QSO.  Instead it appears that SMM\,J02399$-$0136 comprises a merger
between a far-IR-luminous, but highly obscured starburst, the BAL QSO
host and a faint third component. Our findings suggest that this
archetypal SMG and its immediate environment mark a vast and complex
galactic nursery and that detailed studies of other SMGs are likely to
uncover a similarly rich diversity of properties.
\end{abstract}


\section{Introduction}

Submillimetre- (submm-)selected galaxies (SMGs) were discovered using
the Submm Common-User Bolometer Array \citep[SCUBA --][]{holland99},
exploiting gravitational amplication by massive, foreground clusters
\citep{smail97, smail02}.  They are responsible for a significant
fraction of the cosmic submm and far-IR background \citep{fixsen98}
and hence constitute an important class of distant galaxies.
The Very Large Array (VLA) facilitated identification of the first
SMGs, via high-resolution 1.4-GHz imaging \citep{smail00, ivison00}
enabling the acquisition of spectroscopic redshifts, e.g.\ $z=2.8$ for
the first SMG to be discovered, SMM\,J02339$-$0136 \citep[][hereafter
SMM\,J02399 and I98]{ivison98}. \defcitealias{ivison98}{I98} The
rest-frame UV spectral properties of SMM\,J02399 indicated that this
system contained a low-ionisation, BAL quasar, denoted L1 by
\citetalias{ivison98}, with a low-surface-brightness companion
(possibly due to scattered AGN light), L2, associated with a large
Ly$\alpha$-emitting cloud, extending over at least 20\,kpc
(\citetalias{ivison98}; \citealt{ivison99, vm99, vc01}). \citet{vc01}
report an absorption complex at $-$1,000\,km\,s$^{-1}$ and, for
high-ionisation lines, at $-$6,700\,km\,s$^{-1}$, both measured
relative to L1 (at $z=2.795$, which is $\sim 270$\,km\,s$^{-1}$
blueward of L2).  Mid-IR and X-ray data have suggested that the
immense luminosity of this system, $L_{\rm bol}=1.2 \times
10^{13}$\,L$_{\odot}$, arises roughly equally from the AGN and
starburst components \citep{bautz00, lutz05, valiante07}.

SMM\,J02399 was also the first SMG whose molecular gas was detected
via CO \citep{frayer98}.  This exploited the large positive
$K$-correction afforded by even mildly-excited high-$J$ CO lines and
the redshifting of their frequencies to transmissive parts of the
atmosphere \citep[see also][]{brown91, brown92, solomon92,
sd05}. Since then, great progress has been made in discerning the
H$_2$ gas mass distribution in distant SMGs, helping to characterise
their evolutionary state \citep{frayer99, ivison01}; recently, more
sensitive and higher resolution work has employed the Institut de
Radioastronomie Millim\'etrique's (IRAM's) Plateau de Bure
Interferometer \citep[PdBI;][]{downes03, neri03, greve05, kneib05,
tacconi06, tacconi08}. On the basis of CO \jthreetwo\ observations,
\citet{genzel03}, \defcitealias{genzel03}{G03} \!\!\!\!hereafter
\citetalias{genzel03}, argued for the presence of a massive and
rapidly rotating disk in SMM\,J02399, associated with L1, although
they could not rule out an alternative configuration with two galaxies
orbiting one another. They reported a $\sim 3$-arcsec structure in
their sensitive rest-frame 335-$\mu$m continuum imaging, associating
this with L1.

Only a handful of SMG studies \citep{greve03, hainline06, carilli10}
have made use of the CO \jonezero\ line -- the transition most capable
of tracing the bulk of the H$_2$ gas, with its minimal excitation
requirements ($n_{\rm crit}({\rm H}_2)\sim 410$\,cm$^{-3}$ and $\rm
E_{\rm 10}/k_{\rm B} \sim 5.5$\,{\sc k}). Instead, high-$J$ (and thus
high-excitation) CO transitions have usually been used. CO \jthreetwo\
and higher-$J$ rotational levels probe the star-forming molecular gas
\citep[e.g.][]{zhu03, yao03} and thus give a potentially biased view
of the {\it total} gas mass and gas distribution.

It must be noted that even when it comes to the properties of the
star-forming gas, comparisons between SMGs and local ULIRGs are
difficult since the high-$J$ CO lines routinely obtained for SMGs have
only recently become available for the {\it global} molecular gas
reservoirs of ULIRGs \citep[e.g.][]{papadopoulos07c}.  Observations of
the CO \jonezero\ line, on the other hand, allow the most
straightforward estimate of metal-rich H$_2$ gas mass, its
distribution, and the enclosed galactic dynamical mass (assuming
dynamically relaxed systems).  Comparisons with the available CO
\jthreetwo\ also allow an assessment of the global molecular gas
excitation conditions, which are in turn a powerful indicator of the
star-forming and non-star-forming molecular gas fractions in
galaxies. The latter has been found to vary considerably, even among
vigorous starbursts in the local Universe \citep{weiss05a} and at
higher redshifts \citep[e.g.][]{pi02}.

Here, we report sensitive new interferometric observations of
SMM\,J02399, exploiting the arrival of Ka (26.5--40\,GHz) Expanded VLA
(EVLA) receivers at the VLA to trace the bulk of its molecular gas
through its CO \jonezero\ line emission redshifted into a hitherto
inaccessible frequency range \citep[over half of radio-identified SMGs
lie at $z=1.88$--3.35 corresponding to CO \jonezero\ in the Ka
frequency band, ][]{chapman05}.  This well-studied and bright target
provides an obvious starting point for future CO \jonezero\ surveys of
the SMG population with EVLA. We also present sensitive, new 1.4-GHz
continuum imaging, spectroscopic imaging in the rest-frame UV from the
Keck-{\sc i} telescope, and deep multi-filter imaging with {\it
Spitzer} and the Advanced Camera for Surveys (ACS) aboard the {\it
Hubble Space Telescope}.

Throughout this paper we assume a flat cosmological model with
$\Omega_\Lambda=0.7$ and $H_0=70$\,km\,s$^{-1}$\,Mpc$^{-1}$. 

\section{Observations}
\label{observations}

\subsection{CO \jonezero\ observations using the EVLA}
\label{vlaco}

Around 70\,hr of data were obtained with the VLA in its C
configuration during 2009 July--September, around half via
dynamically-scheduled 2-hr blocks, the rest via fixed 6-hr blocks. The
weather conditions were generally excellent. We discarded 20\,hr of
data, taken when a pointing bug was affecting the array, and around
6\,hr of data, taken when the weather was unsuitable.

A compact calibrator, 0239$-$025 (1.0\,Jy at 30.3\,GHz), lies
$\sim1$\,deg from SMM\,J02399, and we used this to track the phase and
amplitude every 3--5\,min, as well as to update the VLA's pointing
model every hour (at 8.4\,GHz). 0137+331 (3C\,48) was used to
calibrate the flux density scale.

Typically, around ten receivers were functional at any one time, each
mounted on an upgraded EVLA antenna. We used continuum mode, giving
dual-polarisation channels, each with a bandwidth of 50\,MHz (or
$\sim500$\,km\,s$^{-1}$ for CO \jonezero\ at $z\sim 2.805$). The
observed frequency of the CO \jonezero\ transition lies below 32\,GHz
for $z\sim 2.8$, where only two intermediate frequencies (IFs), B and
D, can be tuned, so we placed IFs A and C at 32.3\,GHz to constrain
the off-line flux density, even though we expected no significant
continuum emission from either the synchrotron, the free-free or the
dust \citepalias{ivison98}. We alternated the tuning of IFs B and D
between the blue and red peaks seen by \citet{frayer98} and
\citetalias{genzel03}, although some data were taken at non-optimal
frequencies.

The data were processed using standard \AIPS\ procedures developed for
high-frequency observing. Our calibration strategy was able to track
changes in phase and amplitude accurately across the full
0--300\,k$\lambda$ of $uv$ coverage. Imaging was accomplished using
{\sc imagr} within \AIPS.

\subsection{CO \jthreetwo\ and 1.3-mm continuum observations}
\label{pdbi}

\citetalias{genzel03}'s 1.3-mm continuum imaging of SMM\,J02399 from
IRAM's PdBI, and their CO \jthreetwo\ data cube, were kindly made
available to us by L.\ Tacconi and A.\ Baker (private
communication). We use these data here to facilitate the most complete
possible picture of the molecular gas in SMM\,J02399 and its physical
conditions.  The 1.3-mm imaging yields a beam of
1.8-arcsec\,$\times$\,1.4-arcsec (very close to that of our 1.4-GHz
radio map, see \S\ref{20cm}), while the CO \jthreetwo\ map has
significantly lower resolution, 5.2-arcsec\,$\times$\,2.4-arcsec.  A
full description of the observations and their reduction is given in
\citetalias{genzel03}.

\subsection{1.4-GHz imaging with the Very Large Array}
\label{20cm}

Data at 1.4\,GHz were obtained for the Abell\,370 cluster field during
August and September in both 1994 [B configuration] and 1999 [A],
recording $7\times 3.125$-MHz channels in each of two
dual-polarisation IF pairs. 0259$+$077 [B] and 0323$+$055 [A] were
used to track amplitude and phase and 0137$+$331 was used for flux and
bandpass calibration. Data processing and imaging followed the
description given by \citet{owen08}. The resulting continuum map has a
noise level of $\sim5.7$\,$\mu$Jy\,beam$^{-1}$ near SMM\,J02399, with
a 1.8-arcsec\,$\times$\,1.6-arcsec synthesised beam (position angle,
PA, 166\,deg).

\subsection{{\it Hubble Space Telescope} and {\it Spitzer} observations}
\label{hst}

%
%
\begin{figure}
\begin{center}
  \includegraphics[scale=0.287,angle=0]{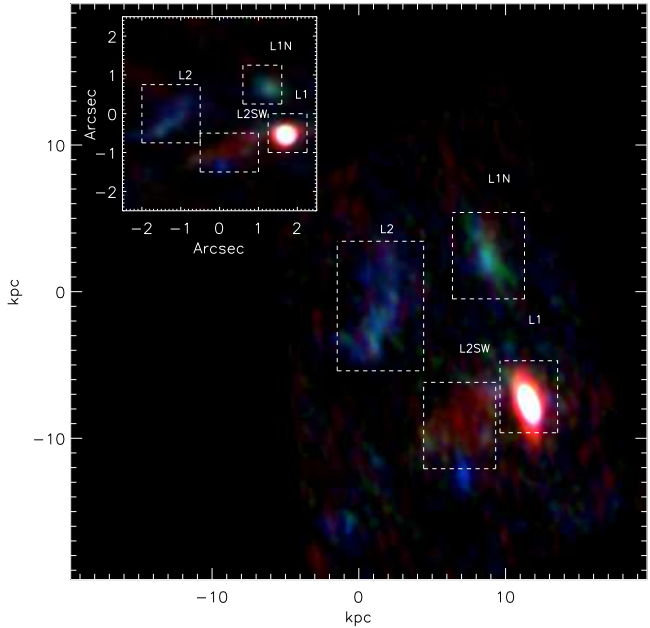}
  \caption{True-colour representations of the {\it HST}/ACS SM4 ERO
    and NICMOS observations of SMM\,J02399 (comprising F475W, F625W
    and F160W exposures) as seen before ({\it inset}) and after ({\it
      main panel}) correcting for the gravitational lensing model
    described in \S\ref{lensmodel}.  In the rest-frame UV imaging, we
    identify the BAL QSO, L1, as well as a marginally extended and
    somewhat bluer companion, L1N, and the much bluer L2.  In
    contrast, the NICMOS image identifies an extended source, L2SW,
    which is very red in the rest-frame UV--optical. This source
    appears to coincide with the far-IR-bright and gas-rich emission
    region within this system. These four primary emission regions are
    marked.}
  \label{model}
\end{center}
\end{figure}

Following the successful Servicing Mission \#4 in 2009 May,
observations were obtained using the Advanced Camera for Surveys (ACS)
during 2009 July (PID 11507), for a total of 6,780, 2,040 and 3,840\,s
in the F475W ($g$), F625W ($r$) and F814W ($i$) bands,
respectively. The data were reduced using the {\sc multidrizzle}
package and the images are available\footnote{\tt
http://archive.stsci.edu/hst/sm4earlypublic}.

In addition, we have retrieved Near Infrared Camera and Multi-Object
Spectrometer (NICMOS) imaging of SMM\,J02399 (PID 11143). These were
reduced and combined in the standard manner from the pipeline-reduced
data in the Space Telescope European Coordinating Facility
archive. They comprise 2,560\,s of integration with the NIC2 camera in
the F110W ($J$) and F160W ($H$) filters.

Finally, Abell\,370 was observed with {\it Spitzer} \citep{werner04}
in all the IR Array Camera \citep[IRAC --][]{fazio04} bands
(3.6--8.0\,$\mu$m) in 2006 February (PID: 137).  The total integration
time was 9\,ks in each of the four bands: 3.6, 4.5, 5.8 and 8\,$\mu$m.

\subsection{Keck spectroscopic observations}
\label{lya}

Keck-{\sc i} long-slit spectroscopy was obtained during 2001 October
using LRIS \citep{oke95} to step through seven positions across the
object, integrating for 2\,ks in each, with a 1-arcsec slit and the
low-resolution 300\,lines\,mm$^{-1}$ grism.  A full description of the
observations and data reduction are presented in \citet{santos04}.

\subsection{Astrometric alignment}

The bright point source, L1, visible in many wavebands, has allowed us
to remove the positional ambiguity that would otherwise hamper a
detailed comparison of the kind undertaken here.

We adopt the VLA 1.4-GHz map as our fundamental coordinate frame and
shift the other frames to this one to align the emission from L1. We
employed local phase calibrators (1.0, 10.6 and 13.0\,deg from
SMM\,J02399), with positions known to $\sim$1\,milliarsec, and
residual phase errors after calibration were small, so we estimate the
absolute alignment of the 1.4-GHz and CO \jonezero\ coordinate frames
should be better than 0.1\,arcsec, leaving aside uncertainties
relating to signal-to-noise and beam size (discussed in
\S\ref{gas}). The astrometry of the {\it HST} images was firstly
calibrated against a wider-field optical image of the cluster which
had in turn been aligned onto the FK5 coordinate system defined by
United States Naval Observatory optical catalogue to an r.m.s.\
precision of $\sim$0.2\,arcsec.  To align the IRAM data (1.3-mm
continuum and CO \jthreetwo) with the 1.4-GHz continuum emission, we
shifted the former to the east by 0.5\,arcsec in R.A., which is
consistent with the claimed astrometric precision of the PdBI
images. To align the optical/near-IR data with the 1.4-GHz continuum
emission, we shifted the former in Dec.\ by +0.15\,arcsec.

\subsection{The lensing model}
\label{lensmodel}

Having described the large collection of new and archival data at our
disposal, we describe now how we deal with the gravitational lensing
experienced by SMM\,J02399. We correct for the lensing distortion and
magnification using a mass model of the Abell\,370 cluster. This is
discussed in detail by \citet{richard09} and is based on ten
multiply-imaged systems detected in the {\it HST}/ACS data
(\S\ref{hst}). We derive a total magnification factor of $2.38\pm
0.08$ at the location of SMM\,J02399 in a direction roughly aligned
with L1--L2 and we use the model to reconstuct the morphology in the
source plane using {\sc lenstool} \citep{kneib96}. The effect of the
correction can be seen in Fig.~\ref{model}.

\section{Analysis and Results}
\label{results}

\subsection{Taxonomy}

We begin with a brief description of the SMM\,J02399 system, as seen
in the rest-frame UV--optical true-colour imaging from {\it HST}.  In
what follows we will refer to four main components seen in
Fig.~\ref{model}: L1 -- the BAL quasar to the west, the brightest UV
emitter in the region; L1N -- the bright, compact source, $\sim$8\,kpc
north of L1; L2 -- the low-surface-brightness feature, $\sim$10\,kpc
east of L1; L2SW, the red feature seen in our NICMOS imaging extending
$\sim$6\,kpc to the south-east of L1.

%
%
\begin{figure*}
\begin{center}
  \includegraphics[scale=0.735,angle=0]{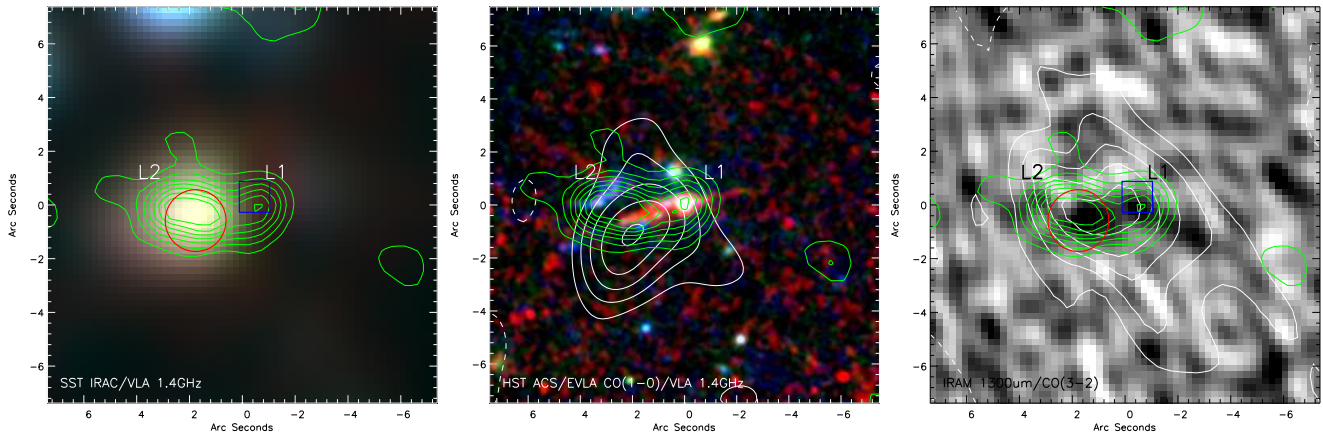}
  \caption{{\it Left:} A true-colour 3.6, 4.5 and 5.8-$\mu$m image of
    SMM\,J02399, where we have removed the contribution from L1 by
    scaling and subtracting a point source from each of the bands,
    revealing emission coincident with L2SW. We overlay on this the
    1.4-GHz continuum contours (green) and the positions of the
    850-$\mu$m and X-ray centroids (red circle --
    \citetalias{ivison98}; blue square -- \citealt{bautz00}, as
    revised by \citetalias{genzel03}). {\it Middle:} CO \jonezero\
    (white contours) from an image made with a Gaussian taper of
    50\,k$\lambda$, plus 1.4-GHz continuum (green), overlaid on the
    {\it HST} imaging. CO contours: $-3, -2, 2, 3, 4, 5,
    6\times\sigma$, where $\sigma=50\,\mu$Jy\,beam$^{-1}$; 1.4-GHz:
    $-4, -2, 2, 4, 6, 8, 10\times\sigma$, where $\sigma =
    5.7$\,$\mu$Jy\,beam$^{-1}$.  {\it Right:} Contours (white) of the
    CO \jthreetwo\ velocity-integrated emission superimposed on a
    greyscale image of the 1.3-mm continuum emission
    \citepalias{genzel03}. \citetalias{genzel03} suggested that L1 is
    the primary repository of SMM\,J02399's dust and gas. It is clear
    that the dusty gas reservoir is not centred on L1, but is
    coincident with L2SW -- the very red source highlighted by the
    NICMOS imaging.}
  \label{morphology}
\end{center}
\end{figure*}

%
%
\begin{figure}
\begin{center}
  \includegraphics[scale=0.34,angle=0]{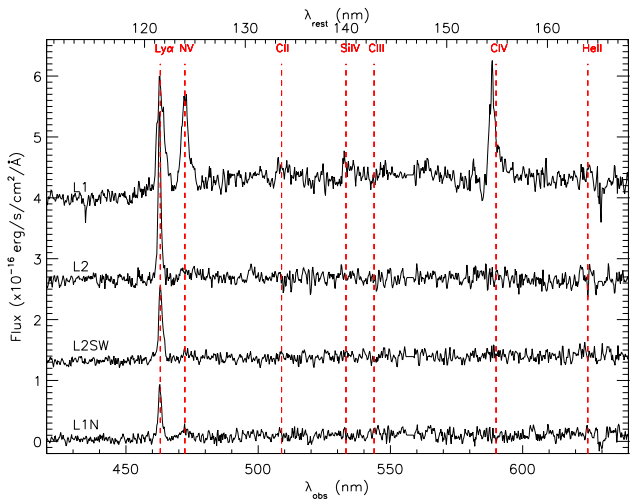}
  \caption{Keck/LRIS spectroscopy of the four regions identified in
    Fig.~\ref{model}. All four regions display strong Ly$\alpha$
    emission and -- with the exception of L1 -- weak or undetected
    rest-frame UV continuum emission.  The spectrum of L1 displays the
    strong, self-absorbed emission lines characteristic of BAL
    QSOs. The three other regions display Ly$\alpha$ and weak N\,{\sc
    v} emission, suggesting local ionisation sources may reside in
    these components. The wavelengths of potential emission features
    are marked.}
  \label{spectra}
\end{center}
\end{figure}

\subsection{Radio continuum morphology}

In Fig.~\ref{morphology} (central panel) we see the 1.4-GHz
synchrotron emission overlaid as green contours on the {\it HST}
true-colour image, together with 50-k$\lambda$-tapered CO contours. A
knot of synchrotron emission is coincident with L1 but the majority of
the radio emission extends to the east, close to the extended near-IR
source, L2SW. It is coincident to within the uncertainties with the
original SCUBA 850-$\mu$m continuum centroid, with the PdBI 1.3-mm
continuum emission, and with the CO \jthreetwo\ line emission
\citepalias{ivison98, genzel03}, as shown in the right-hand panel of
Fig.~\ref{morphology}.

There are two possible explanations for the radio morphology of this
system.  First, the radio morphology could be due to a bipolar jet
from L1 -- a known X-ray-bright BAL quasar -- which encounters only
one working surface: the cloud of molecular gas traced by the CO and
dust emission, to the east of L1. In this scenario, L2 and L2SW may be
shock-excited regions -- L2 lying at the point where the jet
terminates -- with the entire system powered by the kinetic energy of
the jet \citep[e.g.][]{pt05}. Taking this interpretation, SMM\,J02399
is reminiscent of the system described by \citet{jozsa09}, where
extended emission originating at the centre of IC\,2497 at $z=0.05$,
probably a radio jet, points in the direction of Hanny's Voorwerp -- a
region in a neighbouring H\,{\sc i} cloud, first seen because of its
shock-excited, optical emission lines.

Alternatively, and more plausibly, the extended emission may be
associated with a heavily dust-obscured starburst between L1 and L2.
As we shall see in the next section, the host of this starburst, L2SW,
is identified through IRAC and NICMOS rest-frame optical/near-IR
emission, as well as wispy rest-frame UV emission structures
representing leakage from patches of low obscuration (similar to the
scenario proposed for SMM\,J14011$+$0252 by \citealt{ivison01}).  The
spatial coincidence of this source with the CO reservoir, as well as
the 850-$\mu$m and 1.3-mm continuum emission, strongly suggests it is
the seat of the hyperluminous starburst within this system.

As an aside, the lack of emission at 32.3\,GHz rules out the
possibility that the AGN within L1 has a flat-spectrum core.

\subsection{Rest-frame UV/optical/near-IR light}

The most remarkable feature of the 2-D spectroscopy described in
\S\ref{lya} is that Ly$\alpha$ is seen across the entire 1,600-kpc$^2$
region shown in Fig.~\ref{model}. The scale of the Ly$\alpha$ nebular
was hinted at by early data \citepalias{ivison98} but we can see now
that SMM\,J02399 is a classic `Ly$\alpha$ blob'
\citep[e.g.][]{matsuda04}. There is a velocity gradient in Ly$\alpha$
as one moves from L1 to L2. Secondary peaks in Ly$\alpha$ intensity
and velocity dispersion are coincident with L2 -- the local velocity
gradient is around $\sim 500$\,km\,s$^{-1}$. UV continuum is
co-spatial with the line emission over much of the region, but none is
seen near L2.

Fig.~\ref{spectra} shows 1-D spectra of the four regions identified in
Fig.~\ref{model}. Strong Ly$\alpha$ and N\,{\sc v} emission dominate
the spectrum of L1, along with weak He\,{\sc ii} ($\lambda$164.5nm)
\citepalias[e.g.][]{ivison98} at $625.17\pm 0.20$\,nm. L2 and L2SW are
practically indistinguishable, spectroscopically. Regarding the
Ly$\alpha$ emission, and the proximity of an AGN/starburst,
SMM\,J02399 is very similar to the Ly$\alpha$ nebula presented by
\citet{scarlata09}.  L1N, L2 and L2SW all show weak N\,{\sc v}
emission suggesting that they contain ionising sources.

L1 dominates the SMM\,J02399 system in IRAC wavebands. We show in the
left-hand panel of Fig.~\ref{morphology} the true-colour IRAC image
after subtraction of a point source at the position of L1. It is
immediately apparent that L2SW emits strongly at rest-frame
0.95--1.5\,$\mu$m.

Using the optical photometry from \citetalias{ivison98} and the flux
densities of L2 determined from the NICMOS and IRAC imaging
(\S\ref{hst}; see also Figs~\ref{model} and \ref{morphology}), we find
that the spectral energy distribution (SED) is well fit with a
reddened and very young stellar population ($A_V\sim 1.5$;
$\sim$10\,Myr old).  From this we derive a best-fit, rest-frame,
lensing-corrected $H$-band luminosity for the galaxy, $M_H\sim -25.4$,
equivalent to a stellar mass of $\sim$10$^{11}$\,M$_\odot$ using a
light-to-mass ratio appropriate for a young burst ($L_H/M\sim 3$),
with an uncertainty of at least a factor of $3\times$
\citep{borys05}. This confirms the visual impression
(Fig.~\ref{morphology}) that L2 corresponds to a massive stellar
system.  We note that the rest-frame 1.5--2\,$\mu$m SED of L2 is flat,
suggesting that an obscured AGN is not dominating its near-IR
luminosity (cf.\ Hainline et al.\ 2010, in preparation). We stress
that the combined SED of L1 and L2 is dominated by the AGN emission
from L1. Only the combination of gravitational lensing and the
high-resolution imaging used here allows us to decouple the AGN
activity from that of the luminous, obscured starburst within this
system. Similarly detailed studies of other SMGs may also uncover a
rich variety in their properties \citep[e.g.][]{ivison00, ivison08}.

\subsection{Molecular gas}
\label{gas}

%
%
\begin{figure}
\begin{center}
  \includegraphics[scale=0.45,angle=270]{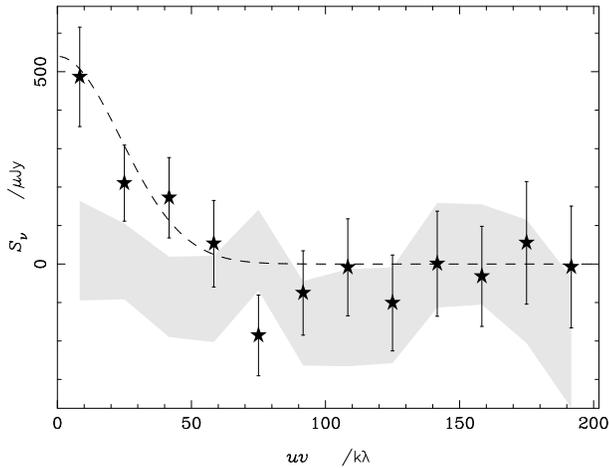}
  \caption{Amplitude of the real visibility components as a function
  of baseline length. The amplitude of the imaginary components, whose
  1-$\sigma$ range is indicated in grey, is consistent with zero. A
  Gaussian consistent with a source size of 10\,k$\lambda$ (20\,arcsec
  {\sc fwhm}) is shown to guide the eye.}
  \label{uv}
\end{center}
\end{figure}

The other major baryonic component in the system is the cold gas. The
CO \jonezero\ images mentioned in \S\ref{vlaco} were made using a
variety of Gaussian $uv$ tapers (125, 75, 50\,k$\lambda$) resulting in
synthesised beam sizes of $\sim$1.7, 2.4 and 3.2\,arcsec and noise
levels of 26--50\,$\mu$Jy\,beam$^{-1}$.  The noise level rises as the
longer baselines are weighted down, but the signal due to CO line
emission increases markedly. This is best illustrated by the
amplitudes of the real and imaginary visibility components
(Fig.~\ref{uv}) which suggest that the CO emission is resolved on
baselines longer than $\sim$10\,k$\lambda$, with a total flux density
of approximately 500\,$\mu$Jy.

The position of the CO \jonezero\ emission ($\rm 02^h 39^m 51.96\pm
0.^s03, -01^{\circ} 36' 59.8\pm 0.5''$ J2000; see
Fig.~\ref{morphology}) is consistent with the emission seen by SCUBA
and PdBI. We cannot rule out a contribution to the CO \jonezero\ flux
from the quasar, but the majority of the emission appears to be
coincident with L2SW.

The peak and total flux densities in the most severely tapered CO
\jonezero\ map are $325\pm 50$\,$\mu$Jy\,beam$^{-1}$ and $540\pm
150$\,$\mu$Jy. The latter value is sensitive to the manner in which
the measurement is made; however, Gaussian fits and a simple integral
over the emission region in the dirty image with {\sc tvstat} give
results consistent with Fig.~\ref{uv}. We adopt 540\,$\mu$Jy as the
total CO \jonezero\ flux density hereafter. Gaussian fits to the dirty
image provide a very rough indication of the size of the emission
region: 2.0\,$\times$\,5.0\,arcsec, with the major axis at PA, $150\pm
20$\,deg, implying a lensing-corrected size of $\sim$25\,kpc. This is
only barely consistent with the $uv$ visibilities (which suggest a
size closer to 10\,arcsec). Without data from the VLA in its D
configuration, we have only loose constraints on the size of the CO
\jonezero\ emitting region and may be underestimating the total flux
density.

To correct for the finite throughput of our IFs, we adopt and adapt
the strategy of \citet{greve03}, correcting upwards by 45 per cent to
account for the breadth of the line seen by \citetalias{genzel03}
relative to our total bandwidth ($\sim$90\,MHz, taking into account
the bandpass rollover), which takes the total flux in CO \jonezero\ to
$S_{\rm 1-0}\,dV = 0.70\pm 0.18$\,Jy\,km\,s$^{-1}$.

As expected, no signal is apparent in the `off-line' IF pair, down to
a 2-$\sigma$ limit of $S_{\rm 32.3GHz}<60\,\mu$Jy. This situation does
not change when tapering the $uv$ data, although the noise doubles
when a 50-k$\lambda$ taper is applied.

The velocity/area-integrated CO brightness temperature in the source
reference frame, $L_{\rm 1-0}= \int_{\Delta V} \int_{A_s} T_{\rm
b}\,dA\,dV$, can be expressed as

\begin{equation}
L_{\rm 1-0} = \frac{3.25\times 10^7}{(1+z)}
\left(\frac{D_{\rm L}}{\nu_{\rm CO}}\right)^2 
\int S_{\rm 1-0}\,dV \, \rm K\,km\,s^{-1}\,pc^2,
\end{equation}

\noindent
where $D_{\rm L}$ is the luminosity distance (Mpc), $\nu_{\rm CO}$ is
the rest-frame frequency (GHz) of the CO \jonezero\ transition and the
velocity-integrated flux density is in Jy\,km\,s$^{-1}$
\citep[e.g.][]{solomon97}.  The corresponding H$_2$ mass, $M({\rm
H}_2)$, is given by $X_{\rm CO} L_{\rm 1-0}$, where $X_{\rm CO}$ is
the CO--H$_2$ luminosity-mass conversion factor. For quiescent
environments, $X_{\rm CO}\sim 5$\,M$_{\odot}$\,({\sc k}\,km\,s$^{-1}$
pc$^2$)$^{-1}$ \citep[e.g.][]{solomon87, sb91}. In extreme starburst
environments, like those prevailing in SMM\,J02399, $X_{\rm CO}\sim
1$\,M$_{\odot}$\,({\sc k}\,km\,s$^{-1}$\,pc$^2$)$^{-1}$ may be more
suitable \citep{ds98}, which corrects semi-empirically for the fact
that the molecular gas reservoir in the dissipative mergers that
typically characterise such starbursts is a continuous medium with a
significant stellar mass component rather than an ensemble of
virialised clouds (so that CO line widths no longer reflect only the
mass of gas). In this case, $M({\rm H}_2)\sim (1.0\pm 0.3) \times
10^{11}$\,M$_{\odot}$ (after correcting for lensing amplification).

An absolute lower limit for the H$_2$ gas mass can be obtained by
assuming that the $^{12}$CO \jonezero\ emission is optically thin. For
local thermodynamic equilibrium,

\begin{eqnarray}
\frac{M({\rm H}_2)}{L_{1-0}} & \sim & 0.08 
\left[\frac{g_1}{Z} \, e^{-T_{\circ}/T_{\rm k}}
\left(\frac{J(T_{\rm k})-J(T_{\rm bg})}{J(T_{\rm k})}\right)\right]^{-1}
\nonumber \\
&& \times \left(\frac{\rm [CO/H_2]}{10^{-4}}\right)^{-1}
\frac{\rm M_{\odot }}{\rm K\,km\,s^{-1}\,pc^2},
\end{eqnarray}

\noindent
where $T_{\circ} = E_1/k_{\rm B}\sim 5.5$\,{\sc k}, $J(T)=T_{\circ}
(e^{T_{\circ}/T}-1)^{-1}$, $T_{\rm bg}=(1+z)\,T_{\rm CMB}\sim
10$\,{\sc k} (the temperature of the cosmic microwave background at
$z=2.8$), $g_1=3$ -- the degeneracy of level $n=1$, $Z\sim 2\,(T_{\rm
k}/T_{\circ})$ -- the partition function, and $\rm [CO/H_2]\sim
10^{-4}$ for a Solar metallicity environment \citep{bs96}.  For
typical star-forming gas, where $T_{\rm k}\sim 40-60$\,{\sc k}, this
yields $\langle X^{\rm thin}_{\rm CO}\rangle \sim 0.65$\,M$_{\odot}$
({\sc k}\,km\,s$^{-1}$ pc$^2$)$^{-1}$, and thus $M({\rm H}_2)_{\rm
min}\sim 6\times 10^{10}$\,M$_{\odot}$ is the minimum plausible
molecular gas mass in this system. Of course, CO traces metal-rich gas
and in metal-poor environments (and/or intense far-UV radiation
fields) more molecular gas may be present where CO has been
dissociated. In this sense, and because we cannot rule out that we
have missed a more extensive CO \jonezero\ component due to the lack
of D-configuration EVLA data, the latter value is a firm lower limit.

The availability of both CO \jonezero\ and \jthreetwo\ allows a
rudimentary probe of the average molecular gas excitation state.  This
can provide powerful clues about whether a galaxy-wide starburst
occurs over most of the available molecular gas reservoir (pushing the
global $r_{3-2}$ ratio up), or whether a smaller gas mass fraction is
involved in fueling the starburst. High global high-$J$/low-$J$ CO
line ratios, indicative of galaxy-wide starbursts involving most of
the ambient gas reservoir, have been observed in various high-redshift
systems \citep[e.g.][]{papadopoulos05, weiss07}, but low ratios in
vigorously starbursting systems have also been detected \citep{pi02,
daddi09, dan09}.

For the reported CO \jthreetwo\ velocity-integrated line flux of
$I_{3-2} =\int S_{\rm 3-2}\,dV = 3.0\pm 0.3$\,Jy\,km\,s$^{-1}$
(\citealt{frayer98}, \citetalias{genzel03}), the globally-average CO
(3$\rightarrow$2)/(1$\rightarrow$0) brightness temperature ratio would
be $r_{\rm 3-2/1-0}=(\nu_{1-0}/\nu_{3-2})^2 (I_{3-2}/I_{1-0}) =
0.48\pm 0.13$. This line ratio is well within the domain expected for
globally star-forming molecular gas, although it lies towards the low
end of the expected range. Quiescent Milky Way clouds have $\rm
r_{3-2/1-0}\sim 0.2-0.3$ whilst the typical average value for
starburst nuclei is $r_{3-2/1-0}=0.65$ \citep{devereux94, yao03}.  A
grid of radiative transfer large-velocity gradient (LVG) models,
constrained by the observed ratio and $T_{\rm k}\geq T_{\rm dust}$
(since gas is always warmer than the concomitant dust where the gas is
heated by far-UV radiation and/or turbulence), with the latter being
$T_{\rm dust}\ga 40$\,{\sc k} \citepalias{ivison98} yields $n({\rm
H}_2)\sim $300--10$^3$\,cm$^{-3}$, for $T_{\rm k} = 40-110$\,{\sc k},
for various values of the average gas velocity gradient $dV/dr$.
Given that other CO line ratios are not available to us, there is an
irreducible $n({\rm H}_2)-T_{\rm k}$ degeneracy (the higher the
temperature, the lower the density reproducing a given CO ratio,
typically).  If we further constrain the LVG solutions by allowing
only those corresponding closely to virialised molecular gas velocity
fields where

\begin{equation}
K_{\rm vir} = \frac{\left(dV/dr\right)_{\rm
    obs}}{\left(dV/dr\right)_{\rm vir}}\sim 1.54\frac{\rm
  [CO/H_2]}{\sqrt{\alpha}\, \Lambda_{\rm CO}}\left[\frac{\langle n({\rm
      H}_2)\rangle}{10^3\,{\rm cm}^{-3}}\right]^{-1/2}\sim 1,
\end{equation}

\noindent
where $K_{\rm vir}\gg 1$ and $K_{\rm vir}< 1$ correspond to unbound
gas motions and unphysical sub-virial motions, respectively, we find
that the solutions with $T_{\rm k}\sim 45-50$\,{\sc k} are mostly
favoured. We note, however, that super-virial solutions are plausible
if the CO \jonezero\ emission originates predominantly from a cold,
extended gas reservoir; indeed, $K_{\rm vir}\gg 1$ solutions occupy
the majority of the $n({\rm H}_2)-T_{\rm k}$ parameter space
compatible with the observed line ratio, typically with $n({\rm
H}_2)\sim 10^3$\,cm$^{-3}$ and $T_{\rm k}\sim 55-90$\,{\sc k},
conditions that often characterise the gas in starbursts
\citep[e.g.][]{aalto95}.

\section{Discussion and concluding remarks}
\label{discussion}

Based on the rest-frame UV and optical imaging, it appears that the
SMM\,J02399 system comprises at least three significant components:
L1, L1N and L2SW. Are these all massive baryonic galaxies within a
very large reservoir of molecular gas, each caught during a different
phase of their development? Or does L1 dominate the system, with L2SW,
L1N (and L2) representing mere outflows and reflections?

\citetalias{genzel03} argued that the dust and gas seen towards
SMM\,J02399 are associated with L1 and that the dynamics of the system
as a whole are best described as a massive disk. Based on the data
described here and elsewhere, there is little doubt that L1 does
represent a significant concentration of mass: it is the brightest
rest-frame UV emitter in the region; it is bright in the rest-frame
0.95--2.1\,$\mu$m waveband and hosts an X-ray-bright BAL quasar.
However, Fig.~\ref{morphology} suggests that \citetalias{genzel03}'s
main assertion may not correct because the mm/submm, radio and CO
emission lie predominantly on and around L2SW, although some weak
emission does appear to be associated with L1.

Moving on to L2 and L2SW, \citet{vc01} and \citetalias{genzel03}
argued that L2 is due to scattered light from L1, although
\citetalias{genzel03} did concede that the L1/L2 system may comprise
two galaxies orbiting one another. As we have argued, the true-colour
{\it Spitzer} imaging presented in Fig.~\ref{morphology} shows that
{\it L2SW is a luminous source at rest-frame 0.95--1.5\,$\mu$m.} We
stress that several of these bands are not expected to suffer strong
contamination by emission lines, hence we are seeing continuum
emission, not scattered light.

We conclude that SMM\,J02399 comprises at least two galaxies --
possibly as many as four -- caught at different points during their
evolution. In the evolutionary scheme outlined by \citet{page04}, L1N
may be the most evolved of these systems, followed by L1, L2 and the
gas- and dust-rich system, L2SW, whose rest-frame far-IR emission led
to the system's discovery. However, this sequence could just as easily
be related to mass: we have either one or two UV systems with modest
stellar mass, the QSO host galaxy, which must be massive according to
\citet{magorrian98}, plus a massive, obscured starburst.

The major result from our study is the discovery of a massive, cold
and apparently extended gas reservoir in SMM\,J02399.  Tentative
evidence for luminous CO \jonezero\ emission in SMGs had been
published previously \citep{greve03, hainline06}.
\citeauthor{greve03} used the VLA to observe ERO\,J16452+4626.4, a
submm-bright starburst at $z=1.4$, finding $r_{2-1/1-0}=0.6\pm 0.2$
and $r_{5-4/1-0}=0.10\pm 0.05$. Similarly, \citeauthor{hainline06}
observed the AGN-dominated $z=3.4$ SMG, SMM\,J13120+4242, using the
Green Bank Telescope, and found a brightness temperature ratio,
$r_{4-3/1-0}=0.26\pm 0.06$, indicative of a massive, sub-thermally
excited molecular gas reservoir. Our new observations suggest that
massive, cold gas components must be common amongst the SMG
population.

In retrospect this discovery shouldn't come as a great surprise.  As
is widely appreciated, optical/IR lines can lead to misleading views
of velocity fields and galaxy types/sizes. They trace ionised gas,
whose distribution traces the star formation and AGN activity within
the galaxy, rather than its underlying structure. Equally, however,
the excitation requirements of high-$J$ CO lines tie them to a
particular star-formation area, so they do not give a faithful picture
of the true H$_2$ mass and dynamical mass distributions in galaxies
either. The compact ($\sim$0.5--1\,kpc) emitting regions thereby
revealed \citep{tacconi08} are unlikely to trace the true dimensions
of galaxies. It can be argued that high-$J$ observations simply return
the star-formation efficiency (SFE) per {\it star-forming} H$_2$ gas
mass, rather than the more fundamental SFE per {\it total} H$_2$ gas
mass. Indeed, previous high-$J$ CO studies have uncovered low
$r_{6-5/3-2}$ and $r_{7-6/3-2}$ brightness temperature ratios
($\sim$0.15--0.35) in several SMGs \citep{tacconi06} and such a low
global high-$J$/low-$J$ CO ratio is exactly how a massive,
low-excitation H$_2$ gas phase would betray itself at high redshifts.

A similar situation exists at $z\sim 0$ where low-excitation H$_2$ gas
is found to extend well beyond compact nuclear starbursts
\citep[e.g.][]{weiss05a} while containing the {\it bulk} of the H$_2$
gas mass. Line ratios comparable to those seen in the SMGs can be
recovered by degrading similar observations of local starbursts and
vigorously star-forming spirals \citep[see][]{papadopoulos98} to the
spatial resolution typically available in CO observations of the
distant Universe (1--10\,kpc).

The confirmation of the presence of significant amounts of cold gas in
SMGs has several significant implications for the interpretation of
this population.  In particular, the large gas reservoir detected in
CO \jonezero\ highlights the danger of drawing far-reaching
conclusions from more biased high-$J$ studies, particularly concerning
the gas mass and gas fraction, the dynamics of SMGs, the evolutionary
links between SMGs and their likely descendants at $z\sim 0$, their
true star-forming potential (i.e.\ their potential present-day mass),
the possible durations of star-forming episodes, and the state of
their velocity fields (ordered and/or disk-like versus the more
chaotic fields typical of mergers) -- {\it all} of these issues are
important, yet the answers remain largely unknown.

In the fullness of time, we expect EVLA, the Atacama Large
Millimetre/submm Array and the {\em Herschel Space Observatory} to
fully compare the excitation properties of gas reservoirs in SMGs,
ULIRGs and QSOs and the relative distributions of star-forming and
quiescent molecular gas, relating this to the activity in these
systems, and their evolution, thereby constraining theoretical models
of galaxy formation \citep{swinbank08}.

\section*{Acknowledgements}

We thank Linda Tacconi and Andrew Baker for generously providing their
IRAM PdBI data, Richard Ellis for granting us access to the LRIS
observations and Julie Wardlow for help with the stellar population
analysis.  IRS acknowledges support from the UK Science and Technology
Facilities Council. AMS and JR gratefully acknowledge a Royal
Astronomical Society Sir Norman Lockyer Fellowship and a Marie Curie
fellowship, respectively.

\bibliographystyle{mnras}
\bibliography{20100107}

\bsp

\end{document}